\documentclass[12pt]{article}
\usepackage{amsmath, amssymb}
\usepackage{graphicx}
\usepackage{natbib}
\usepackage{url} 

\usepackage{xcolor}

\renewcommand{\baselinestretch}{1.5}

\begin{document}

\def\spacingset#1{\renewcommand{\baselinestretch}%
{#1}\small\normalsize} \spacingset{1}

\newtheorem{lem}{Lemma}
\newtheorem{cor}{Corollary}
\newtheorem{prp}{Proposition}


  \title{\bf On non-stationarity of the Poisson gamma state space models}
  \author{Kaoru Irie\\
    Faculty of Economics, University of Tokyo \\ 
    \url{irie@e.u-tokyo.ac.jp} \\
    and \\
    Tevfik Aktekin\\
    Paul College of Business and Economics, University of New Hampshire}
    \date{}
  \maketitle

\bigskip
\begin{abstract}
The Poisson-gamma state space (PGSS) models have been utilized in the analysis of non-negative integer-valued time series to sequentially obtain closed form filtering and predictive densities. In this study, we show the underlying mechanics and non-stationary properties of multi-step ahead predictive distributions for the PGSS family of models. By exploiting the non-stationary structure of the PGSS model, we establish that the predictive mean remains constant while the predictive variance diverges with the forecast horizon, a property also found in Gaussian random walk models. We show that, in the long run, the predictive distribution converges to a zero-degenerated distribution, such that both point and interval forecasts eventually converge towards zero. In doing so, we comment on the effect of hyper-parameters and the discount factor on the long-run behavior of the forecasts. 
\end{abstract}

\noindent%
\vfill

\spacingset{1.5} 

\section{Introduction}

The Poisson-gamma state space (PGSS) model is designed for non-negative integer valued time series data, or the time series of counts \citep{bather1965invariant,smith1986non,harvey1989timea,harvey1989timeb}. Variants of the PGSS model also appear in \cite{gamerman2013non} and \cite{aktekin2012bayesian}. For a count-valued observation at time $t$, or $y_t$, the PGSS model assumes the Poisson sampling distribution's mean, $\theta_t$, evolves via the multiplicative random walk model: 
\begin{equation} \label{eq:model}
\begin{split}
    y_t | \theta _t &\sim \mathrm{Possion}(\theta_t) \\
	\theta _t &= \frac{\theta _{t-1}\eta _t}{\gamma}, \qquad \eta_t \sim \mathrm{Beta}(\gamma a_{t-1},(1-\gamma)a_{t-1} ),
\end{split}
\end{equation}
with $\theta _0 \sim \mathrm{Gamma}(a_0,b_0)$, where $a_0>0$, $b_0>0$ and $\gamma \in (0,1)$ are constant and $a_t$ is a summary statistics and updated via $a_t=\gamma a_{t-1}+y_t$. 
The PGSS model is conjugate, in the sense that the filtering posteriors are computed analytically; $p(\theta _t | y_{1:t})$ is shown to be a gamma distribution. 
See Section~2, as well as \cite{aktekin2013assessment}, for details. 

Although the model’s flexibility and fitting capabilities are limited, the PGSS model has proven to be attractive for its computational efficiency, particularly in real-time monitoring and forecasting settings where data are streaming and posterior updates and fast predictions are necessary. 
Examples of the application of the PGSS models include the analysis and forecast of mortgage default \citep{aktekin2013assessment} and record of accesses to websites \citep{chen2018scalable,irie2022sequential} among others. Extensions to multivariate responses \citep{aktekin2018sequential} and a wider class of sampling distributions \citep{aktekin2020family} has also been discussed. 

Under the PGSS model, the time series of $y_t$ is non-stationary. As is the case for other non-stationary time series, the use of the PGSS model is recommended for only short-term and sequential predictions. For long-term predictions under the PGSS model, the multi-step-ahead predictive distribution is expected to be highly diffuse, rendering it weakly informative and of limited practical use. The purpose of this article is to study the mathematical properties of the the long term behavior of the predictive distribution, a topic that has received little to no attention in the Poisson state space literature. 

\begin{figure}[htbp!]
\centering
\includegraphics[width=\linewidth]{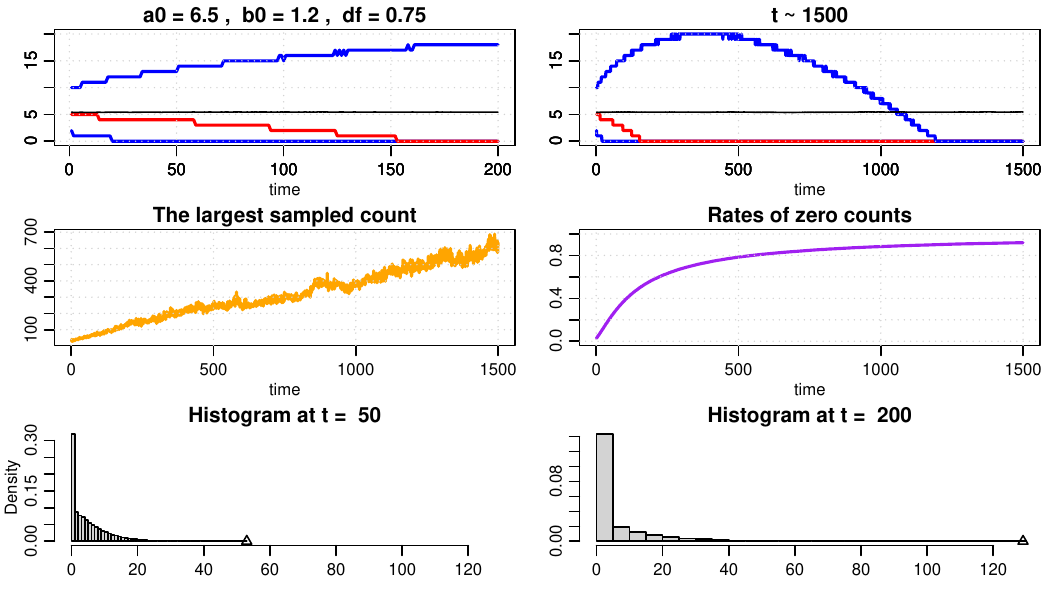} 
\caption{Marginal distribution of $p(y_t)$: Top-left and right: Predictive means (thin, black), median (red) and 10\% and 90\% quantiles (blue); Middle-left: Maximum $y_t$ over $t$; Middle-right: Zero counts rate; Bottom: $y_t$'s histogram at $t=50$ (left) and $t=200$ (right), the triangle symbol indicates the largest sampled value.}
\label{fig:ex}
\end{figure}

To observe the behavior of the multiple-step ahead predictive distribution under the PGSS models, we first sampled the marginal distribution of $y_t$, which is viewed as the $t$-step ahead predictive distribution at time $0$ using the Monte Carlo method. For illustration purposes, we set $a_0=6.5$, $b_0=1.2$ and $\gamma = 0.75$, with a Monte Carlo size of $50000$. The top panels in Figure~\ref{fig:ex} show that the predictive mean is constant, but the predictive quantiles decrease toward zero as $t$ increases. 
In the time series analysis of count data, the predictive median is a popular choice as the point forecast minimizing the mean absolute distance loss. When interpreting these initial numerical results, users should be mindful of the model’s non-stationarity and should not perceive the decline in the predictive median as a reliable future forecast. In a similar vein, we can also obverse that the upper 10\% predictive quantiles converge towards zero, which should not be interpreted as indicating a more accurate or less uncertain future forecast. 

As inferred from the fact that the quantiles are converging towards zero, the probability of a zero count is converging to unity, as empirically observed in the middle-right panel of Figure~\ref{fig:ex}. In contrast, the largest count sampled is increasing as seen in the middle-left panel, suggesting the divergence of the variance of $y_t$. The bottom panels show the histogram of $y_t$ at $t=50$ and $t=200$, summarizing our observations on the behavior of $p(y_t)$. That is, as $t\to \infty$, we have a larger probability mass on $\{  y_t = 0 \}$, while the rest of the probability is assigned to extreme values in the tail. 
These properties are unique in the PGSS model and are not observed in the Gaussian random walk. 

Similar patterns for $p(y_t)$ can also be observed for arbitrary choices of $(a_0,b_0)$ and $\gamma$. Therefore, we conjecture that; 
$\mathbb{E}[y_t]=a_0/b_0$ for all $t$, 
$\mathbb{V}[y_t]\to\infty$ as $t\to\infty$ (Proposition~\ref{prp:moment}), and 
$\mathbb{P}[y_t=0] \to 1$ as $t\to \infty$ (Proposition~\ref{prp:conv}) 
for any choice of hyper-parameters and prove these statements mathematically in this article. To the extent of our knowledge, this is the first attempt to study the multiple-step predictive distribution under the PGSS models analytically, partly because the multiple-step ahead predictive distribution has been evaluated by the Monte Carlo method in practice. Our new theoretical results on the PGSS model reinforce the classical warning against interpreting long-term predictive distributions from non-stationary time series models.

In the rest of this article, we review the definition and known properties of the PGSS model in Section~2, then prove the results on predictive moments and convergence in distribution in Sections 3 and 4, respectively. 

\section{Preliminary results}

\subsection{The PGSS models}
We first specify the components of the PGSS model in detail. The sampling distribution of the PGSS models assumes conditional independence as in  
\begin{equation*}
    p(y_t | y_{1:(t-1)}, \theta _{0:T} ) = p(y_t|\theta _t ) \qquad \mathrm{for \ all } \ t,
\end{equation*}
and each observation follows the Poisson distribution with mean $\theta_t$ whose probability function is given by,
\begin{equation*}
    \mathbb{P}[y_t=y|\theta _t] = \frac{\theta_t^y}{y!} e^{-\theta_t} , \qquad (y=0,1,2,\dots ).
\end{equation*}
In modeling the evolution of the state variable, we also assume conditional independence, 
\begin{equation*}
    p( \theta _t | \theta _{0:(t-1)}, y_{1:(t-1)} ) = p( \theta _t | \theta _{t-1}, y_{1:(t-1)} ) , \qquad  \mathrm{for \ all } \ t,
\end{equation*}
and define the right hand side via the distributional equation given in (\ref{eq:model}), where $\gamma \in (0,1)$ is a pre-specified constant and is usually referred to as a discount factor. 
The model is completed with the initialization of the state variable as $\theta _0 \sim \mathrm{Gamma}(a_0,b_0)$, a gamma distribution with shape $a_0>0$ and rate $b_0>0$ (mean $a_0/b_0$).

\subsection{Known results}
The forward filter is available analytically under the family of PGSS models. If we assume the following:  
\begin{itemize}
    \item Posterior at $t-1$: $\theta _{t-1} | y_{1:(t-1)} \sim \mathrm{Gamma}(a_{t-1},b_{t-1})$, 
\end{itemize}
which is true at $t=0$, then we have 
\begin{itemize}
    \item Prior at $t$:  $\theta _{t} | y_{1:(t-1)} \sim \mathrm{Gamma}(\gamma a_{t-1}, \gamma b_{t-1})$.

    \item Posterior at $t$: $\theta _{t} | y_{1:t} \sim \mathrm{Gamma}(a_{t},b_{t})$, where $(a_t,b_t)$ is recursively updated by 
    \begin{equation*}
        a_t = \gamma a_{t-1} + y_t, \qquad \mathrm{and}\qquad b_t = \gamma b_{t-1} + 1,
    \end{equation*}
    for all $t\ge 1$. Summary statistics $a_t$ is the function of $y_{1:t}$, but $b_t$ is not; 
    \begin{equation*}
    \begin{split}
        b_t = \gamma^t b_0 + \gamma^{t-1} + \cdots \gamma + 1 = \frac{1-\{ 1-(1-\gamma)b_0 \}\gamma^t}{1-\gamma}. 
    \end{split}
    \end{equation*}
    Let $b^{\ast}\equiv 1/(1-\gamma)$. If $b_0 < b^{\ast}$, then $b_{t-1} < b_t < b^{\ast}$ for all $t$ and $b_t\uparrow b^{\ast}$. Conversely, if $b^{\ast}<b_0$, then $b^{\ast} < b_t < b_{t-1}$ for all $t$ and $b_t\downarrow b^{\ast}$. If $b_0=b^{\ast}$, then $b_t=b^{\ast}$ for all $t$. 

    \item The one-step ahead predictive distribution, or $p(y_t|y_{1:(t-1)})$, is a negative binomial distribution. 
    
\end{itemize}

\section{Predictive mean and variance}

Suppose that we are at time $0$ and are interested in the $t$-step ahead marginal predictive distribution, or $p(y_t)$, which is unconditional on $y_{1:(t-1)}$ and $\theta _{1:t}$. The marginal moment of $y_t$ is computed easily by using the Tower Property. 

\begin{prp} \label{prp:moment}
    For any initial states $(a_0,b_0)$ and discount factor $\gamma$, we have 
\begin{equation*}
         \mathbb{E}[ y_t ] = \frac{a_0}{b_0} \quad \mathrm{for \ any \ } t, \quad \mathrm{and}, \quad \mathbb{V}[ y_t ] \to \infty \quad \mathrm{as}\quad t\to\infty. 
\end{equation*}
\end{prp}

\noindent 
{\it Proof.} The one-step ahead predictive mean at $t-1$ is 
\begin{equation*}
    \mathbb{E}[ y_t | y_{1:(t-1)} ] = \mathbb{E}[ \theta _t | y_{1:(t-1)} ] = \frac{a_{t-1}}{b_{t-1}},
\end{equation*}
where the second expectation is taken with respect to $\theta _t | y_{1:(t-1)} \sim \mathrm{Gamma}(\gamma a_{t-1},\gamma b_{t-1})$. Rewriting $(a_{t-1},b_{t-1})$ recursively through $(a_{t-2},b_{t-2})$ and $y_{t-1}$, we obtain  
\begin{equation*}
    \mathbb{E}[ y_t | y_{1:(t-1)} ] = \frac{\gamma b_{t-2}}{\gamma b_{t-2}+1} \cdot \frac{a_{t-2}}{b_{t-2}} +  \frac{y_{t-1}}{\gamma b_{t-2}+1}.  
\end{equation*}
Noting that $a_{t-2}/b_{t-2} = \mathbb{E}[ y_{t-1} | y_{1:(t-2)} ]$, we can take the expectation of both sides with respect to $p(y_{t-1}|y_{1:(t-2)})$ to show $\mathbb{E}[ y_t | y_{1:(t-2)} ] = \mathbb{E}[ y_{t-1} | y_{1:(t-2)} ]$. By repeating this computation, we conclude that $\mathbb{E}[ y_t ] = \mathbb{E}[ y_1 ] = a_0 / b_0$ for all $t$. 

Similarly, the predictive variance can be obtained via 
\begin{equation*}
    \mathbb{V}[ y_t ] = \mathbb{V}[ \mathbb{E}[y_t|y_{1:(t-1)}] ] + \mathbb{E}[ \mathbb{V}[y_t|y_{1:(t-1)}] ] = \frac{1}{b_{t-1}^2} \mathbb{V}[a_{t-1}] + \left( \frac{b_{t}}{\gamma b_{t-1}} \right) \frac{a_0}{b_0}.
\end{equation*}
Noting that $b_t$ and $b_{t-1}$ converge to a non-zero constant, it is sufficient to show the divergence of $\mathbb{V}[a_{t-1}]$. The variance of interest would have the following form: 
\begin{equation*}
    \mathbb{V}[ a_{t-1} ] =  \left( \frac{b_{t-1}}{b_{t-2}} \right) ^2 \mathbb{V}[ a_{t-2} ] + \frac{1}{\gamma}\left( \frac{b_{t-1}}{b_{t-2}} \right) \frac{a_0}{b_0}.
\end{equation*}
Since $b_{t-1}b_{t-2}\to 1/(1-\gamma)^2$ as $t\to\infty$, the sequence of $\mathbb{V}[ a_{t-1} ] / b_{t-1}^2 \to \infty$ as $t\to\infty$. $\square$

\section{Convergence in distribution}

\subsection{Recurrence relation of probabilities of zero count} \label{sec:rec}

Denote the probability generating function (p.g.f.) of the $t$-step ahead predictive distribution, or the marginal distribution of $y_t$, with initialization $\theta _0 \sim \mathrm{Gamma}(a_0,b_0)$ by $\varphi_t(s|a_0,b_0) \equiv \mathbb{E}[s^{y_t}]$. For $t=1$, the predictive distribution is negative binomial:
\begin{equation*}
	\varphi_1(s|a_0,b_0) = \left( \frac{\gamma b_0}{\gamma b_0+1-s} \right) ^{\gamma a_0},
\end{equation*}
for all $|s|<1$. It is worth noting that $\varphi_1(s|a_0,b_0) = \varphi_1(s|1,b_0)^{a_0}$; the shape parameter is the exponent in the p.g.f. We first prove that this is also true for any $t\ge 1$. 

\begin{lem}
    $\varphi_t(s|a_0,b_0) = \varphi_t(s|1,b_0) ^{a_0}$ for any $|s|<1$, $(a_0,b_0)$ and $\gamma$. 
\end{lem}

\noindent
{\it Proof.} We prove this statement by induction. It has already been proven for $t=1$. Assuming $\varphi_{t-1}(s|a_0,b_0) = \varphi_{t-1}(s|1,b_0) ^{a_0}$, we compute $\varphi_t(s|a_0,b_0)$. Note that $\mathbb{E}[ s^{y_t} | y_1 ]$ is viewed as the $(t-1)$-step ahead predictive p.g.f.~with initialization $\theta _1 | y_1 \sim \mathrm{Gamma}(a_1,b_1)$, hence it equals $\varphi_{t-1}(s|a_1,b_1)$. By the Tower Property, we compute $\varphi_t(s|a_0,b_0)$ as 
\begin{equation*}
	\begin{split}
		\varphi_t(s|a_0,b_0) = \mathbb{E}[ \varphi _{t-1}(s|1,b_1)^{a_1} ] &= \varphi _{t-1}(s|1,b_1)^{\gamma a_0} \varphi_1 ( \ \varphi _{t-1}(s|1,b_1) \ |a_0,b_0 ) \\
		&= \left[ \varphi _{t-1}(s|1,b_1) \frac{\gamma b_0}{\gamma b_0 + 1 - \varphi _{t-1}(s|1,b_1) } \right] ^{\gamma a_0},
	\end{split}
\end{equation*}
for any $a_0>0$, including $a_0=1$. This shows $\varphi_t(s|a_0,b_0) = \varphi_t(s|1,b_0) ^{a_0}$. $\square$

\ 

The probability of zero count is obtained by evaluating the p.g.f.~at $s=0$. That is, $\mathbb{P}[y_t=0 | a_0, b_0 ] = \varphi_t(0|a_0,b_0)$. We focus on the case with $a_0=1$ by writing 
\begin{equation*}
	p_t(b_0) \equiv \varphi _t(0|1,b_0) = \mathbb{P}[y_t=0 | a_0=1, b_0 ],
\end{equation*}
because we have $\mathbb{P}[y_t=0 | a_0, b_0 ] = p_t(b_0)^{a_0}$ and, if $p_t(b_0) \to 1$ as $t\to \infty$, then we have $\mathbb{P}[y_t=0 | a_0, b_0 ] \to 1$ for any $a_0>0$. The recurrence equation for the p.g.f.~is translated into that of the probability of zero count as 
\begin{equation} \label{eq:pgf}
		p_t(b_0) = \left[ p_{t-1}(b_1) \frac{\gamma b_0}{\gamma b_0 + 1 - p_{t-1}(b_1) } \right] ^{\gamma},
\end{equation}
where $b_1 = \gamma b_0 + 1$. This probability, $p_t(b_0)$, has several properties as a function of $t$ and $b_0$, which are necessary in the following proof and summarized below.

\begin{lem} \label{lem:tools}
    Fix $\gamma \in (0,1)$ to an arbitrary value. Then, (i) $p_t(b)$ is increasing in $b$ for any fixed $t$, and (ii) $p_t(b)$ is increasing in $t$ for any fixed $b>0$. 
\end{lem}

\noindent
{\it Proof.} (i) It is obvious that $p_1(b) = \{ \gamma b / (\gamma b + 1) \} ^{\gamma}$ is increasing in $b$. Assuming that $p_{t-1}(b)$ is increasing, we rewrite (\ref{eq:pgf}) as,
\begin{equation*}
	p_t(b) = \left[ p_{t-1}(\gamma b + 1) \left( 1 + \frac{1-p_{t-1}(\gamma b + 1)}{\gamma b} \right) ^{-1} \right]^{\gamma},
\end{equation*}
where $p_{t-1}(\gamma b + 1)$ is increasing, and $\frac{1-p_{t-1}(\gamma b+1)}{\gamma b}$ is decreasing, hence the right hand side is increasing in $b$. 

(ii) In this proof, we use $b_0$ and  $b_1=\gamma b_0+1$ instead of $b$ and $\gamma b + 1$ for notational clarity. For $t=1$, we compute the ratio of $p_2(b_0)$ and $p_1(b_0)$ and show that 
\begin{equation*}
    \left( \frac{ p_2(b_0) }{ p_1(b_0) } \right) ^{1/\gamma} = \frac{p_1(b_1) b_1}{ b_1 - p_1(b_1) } \ge 1 \qquad \Leftrightarrow \qquad \left( \frac{\gamma b_1}{\gamma b_1+1} \right) ^{\gamma} \ge \frac{b_1}{b_1+1}.
\end{equation*}
Since the inequality in the right holds for any $b_0$ and $\gamma$, we conclude that $p_2(b_0)\ge p_1(b_0)$ for any $b_0$. Next, assume that $p_t(b_0)/p_{t-1}(b_0) \ge 1$ for any $b_0$. Then, 
\begin{equation*}
    \frac{ p_{t+1}(b_0) }{ p_t(b_0) } = \left[ \frac{ p_{t}(b_1) }{ p_{t-1}(b_1) } \cdot \frac{\gamma b_0+1 -p_{t-1}(b_1)}{\gamma b_0+1 -p_t(b_1) }  \right]^{\gamma} \ge \left[ \frac{ p_{t}(b_1) }{ p_{t-1}(b_1) }  \right]^{\gamma} \ge 1,
\end{equation*}
showing the monotonicity at $t+1$. $\square$

\subsection{Special case: $b_0 = b^{\ast} = 1/(1-\gamma)$}

We first prove the convergence of interest for a particular choice of the initial value: $b^{\ast}=1/(1-\gamma)$. With $b_0 = b^{\ast}$, the updating rule of summary statistics implies $b_t = b^{\ast}$ for all $t$. This property simplifies the computation of recurrence relation (\ref{eq:pgf}). 

\begin{lem} \label{lem:special}
    For any $\gamma \in (0,1)$, $p_t(b^{\ast})\to 1$ as $t\to \infty$. 
\end{lem}

\noindent 
{\it Proof.} Write $p_t \equiv p_t(b^{\ast})$ for any $t$. Note that $p_1 =  \gamma ^{\gamma}$. By using recurrence relation (\ref{eq:pgf}), we obtain this bound as 
\begin{equation*}
		p_t = \left[ p_{t-1} \frac{\gamma b^{\ast} }{\gamma b^{\ast} + 1 - p_{t-1} } \right]^{\gamma} \ge p_{t-1}^{\gamma} \gamma^{\gamma} \ge \gamma^{\frac{\gamma}{1-\gamma} (1-\gamma^t) } \ge \gamma ^{\gamma / (1-\gamma)} > 0.
\end{equation*}
Since $p_t$ is increasing by Lemma~\ref{lem:tools} (ii) and $p_t \in [c,1]$ with $c=\gamma ^{\gamma / (1-\gamma)}$ for all $t$, there exists $\displaystyle p _{\infty} \equiv \lim _{t\to\infty }p_t \in [c,1]$. Then, by taking the limits of both sides of (\ref{eq:pgf}), we observe that 
\begin{equation*}
	\left[ p \frac{\gamma b^{\ast} }{\gamma b^{\ast} + 1 - p} \right]^{\gamma} = p,
\end{equation*}
must hold at $p=p_{\infty}$. For $p \in [c,1)$, the left-hand-side is strictly larger than the right. Hence $p_{\infty}=1$ is the unique solution of the equation above. $\square$

\subsection{Extension to general $b_0$}

Based on Lemma~\ref{lem:special}, the convergence of interest is shown for any $b_0>0$. 

\begin{prp} \label{prp:conv}
    For any $\gamma \in (0,1)$ and $b_0 > 0$, $p_t(b_0)\to 1$ as $t\to \infty$. 
\end{prp}

\noindent 
{\it Proof.} The statement holds for $b_0=b^{\ast}$ by Lemma~\ref{lem:special}. If $b_0 > b^{\ast}$, then by monotonicity in $b_0$ (Lemma~\ref{lem:tools} (i)), we have $p_t (b_0) \ge p_t(b^{\ast})$ for all $t$, which shows $p_t (b_0) \to 1$ as $t\to \infty$. Below, we focus on the case of $b_0 < b^{\ast}$. 

We first show that the effect of initial value $b_0$ diminishes as $t\to\infty$. For any $t_0 \ge 1$, 
\begin{equation*}
	\varphi_{t+t_0}(s|b_0) = \mathbb{E}[ \mathbb{E}[s^{y_{t+t_0}}|y_{1:t_0}] ] = \mathbb{E}[ \varphi_t(s|a_{t_0},b_{t_0}) ], 
\end{equation*}
holds.  
Note that $b_{t_0}$ is arbitrarily close to $b^{\ast}=1/(1-\gamma)$ if $t_0$ is sufficiently large. By setting $s=0$, we express the probabilities of zero count as $p_{t+t_0}(b_0) = \mathbb{E}[ p_{t}(b_{t_0})^{a_{t_0}} ]$. Thus, if $p_{t}(b_{t_0}) \to 1$ as $t\to \infty$, then by the bounded convergence theorem, we have $p_{t+t_0}(b_{0}) \to 1$. By this observation, we are allowed to consider $b_0$ which is sufficiently close to $b^{\ast}$. 

Fix an arbitrary $\delta > 0$. By Lemma~\ref{lem:special}, we find $T>0$ such that $p_t(b^{\ast})>1-\delta /2$ for any $t\ge T$. With this $T$, we use Lemma~\ref{lem:tools} (i) to find an $\varepsilon > 0$ such that $p_T(b^{\ast}) - p_T(b_0) < \delta /2$ for any $b_0 \in ( (1-\varepsilon) b^{\ast},b^{\ast})$. Thus, by using Lemma~\ref{lem:tools} (ii), we have 
\begin{equation*}
    p_t(b_0) \ge p_T(b_0) = [p_T(b_0)  - p_T(b^{\ast})] + p_T(b^{\ast}) \ge (-\delta/2) + (1-\delta/2) = 1-\delta, 
\end{equation*}
which shows the convergence of $p_t(b_0)$ and completes the proof. $\square$

\section*{Acknowledgment}

The authors thank Fumiya Akashi, Takeru Matsuda and Shintaro Hashimoto for discussions. Research of the first author is partially supported by JSPS KAKENHI grant numbers 22K13374. 

\spacingset{0.8}

\bibliographystyle{chicago}
\bibliography{pgss}

@article{gamerman2013non,
  title={A non-Gaussian family of state-space models with exact marginal likelihood},
  author={Gamerman, Dani and dos Santos, Thiago Rezende and Franco, Glaura C},
  journal={Journal of Time Series Analysis},
  volume={34},
  number={6},
  pages={625--645},
  year={2013},
  publisher={Wiley Online Library}
}

@article{aktekin2013assessment,
  title={Assessment of mortgage default risk via {B}ayesian state space models},
  author={Aktekin, Tevfik and Soyer, Refik and Xu, Feng},
  journal={The Annals of Applied Statistics},
  pages={1450--1473},
  year={2013},
  publisher={JSTOR}
}

@article{aktekin2012bayesian,
  title={Bayesian analysis of queues with impatient customers: Applications to call centers},
  author={Aktekin, Tevfik and Soyer, Refik},
  journal={Naval Research Logistics (NRL)},
  volume={59},
  number={6},
  pages={441--456},
  year={2012},
  publisher={Wiley Online Library}
}

@article{aktekin2018sequential,
  title={Sequential {B}ayesian Analysis of Multivariate Count Data},
  author={Aktekin, Tevfik and Polson, Nicholas G and Soyer, Refik},
  journal={Bayesian Analysis},
  volume={13},
  number={2},
  pages={385--409},
  year={2018}
}

@article{aktekin2020family,
  title={A family of multivariate non-{G}aussian time series models},
  author={Aktekin, Tevfik and Polson, Nicholas G and Soyer, Refik},
  journal={Journal of Time Series Analysis},
  volume={41},
  number={5},
  pages={691--721},
  year={2020},
  publisher={Wiley Online Library}
}

@article{bather1965invariant,
  title={Invariant conditional distributions},
  author={Bather, J A},
  journal={The Annals of Mathematical Statistics},
  volume={36},
  number={3},
  pages={829--846},
  year={1965},
  publisher={JSTOR}
}

@article{chen2018scalable,
  title={Scalable {B}ayesian modeling, monitoring, and analysis of dynamic network flow data},
  author={Chen, Xi and Irie, Kaoru and Banks, David and Haslinger, Robert and Thomas, Jewell and West, Mike},
  journal={Journal of the American Statistical Association},
  volume={113},
  number={522},
  pages={519--533},
  year={2018},
  publisher={Taylor \& Francis}
}

@article{harvey1989timea,
  title={Time series models for insurance claims},
  author={Harvey, Andrew C and Fernandes, Clara},
  journal={Journal of the Institute of Actuaries},
  volume={116},
  number={3},
  pages={513--528},
  year={1989},
  publisher={Cambridge University Press}
}

@article{harvey1989timeb,
  title={Time series models for count or qualitative observations},
  author={Harvey, Andrew C and Fernandes, Clara},
  journal={Journal of Business \& Economic Statistics},
  volume={7},
  number={4},
  pages={407--417},
  year={1989},
  publisher={Taylor \& Francis}
}

@article{irie2022sequential,
  title={Sequential modeling, monitoring, and forecasting of streaming web traffic data},
  author={Irie, Kaoru and Glynn, Chris and Aktekin, Tevfik},
  journal={The Annals of Applied Statistics},
  volume={16},
  number={1},
  pages={300--325},
  year={2022},
  publisher={Institute of Mathematical Statistics}
}

@article{smith1986non,
  title={A non-{G}aussian state space model and application to prediction of records},
  author={Smith, R L and Miller, J E},
  journal={Journal of the Royal Statistical Society: Series B (Methodological)},
  volume={48},
  number={1},
  pages={79--88},
  year={1986},
  publisher={Wiley Online Library}
}

\end{document}